\documentclass[10pt,twocolumn,amssymb]{revtex4}
\usepackage{latexsym}
\usepackage{graphics}
\usepackage{amsmath,amsfonts}

\def\NO{\nonumber}

\newcommand{\be}{\begin{equation}}
\newcommand{\ee}{\end{equation}}

\def\bea{\begin{eqnarray}}
\def\eea{\end{eqnarray}}

\def\beqx{\begin{displaymath}}
\def\eeqx{\end{displaymath}}

\newcommand{\bmat}{\left(\begin{array}}
\newcommand{\emat}{\end{array}\right)}

\def\half{\frac{1}{2}}

\def\a{\alpha}

\def\d{\delta}
\def\e{\epsilon}
\def\f{\phi}

\def\k{\kappa}
\def\l{\lambda}
\def\m{\mu}
\def\n{\nu}

\def\p{\pi}

    \def\th{\theta}

\def\s{\sigma}

\def\D{\Delta}

\def\G{\Gamma}

\def\L{\Lambda}

    \def\Om{\Omega}

\def\vf{\varphi}

\def\cc{{\cal C}}

\def\cn{{\cal N}}
\def\co{{\cal O}}
\def\cp{{\cal P}}

\def\cw{{\cal W}}

\def\bo{{\raise-.3ex\hbox{\large$\Box$}}}               
\def\pa{\partial}                                       
\def\face{{\raise.2ex\hbox{$\displaystyle \bigodot$}\mskip-2.2mu \llap {$\ddot
        \smile$}}}                                   
\def\>{\rangle}                                      
\def\<{\langle}                                      

\newcommand{\sub}[1]{\phantom{}_{(#1)}\phantom{}}    
\def\leftrightarrowfill{$\mathsurround=0pt \mathord\leftarrow \mkern-6mu
        \cleaders\hbox{$\mkern-2mu \mathord- \mkern-2mu$}\hfill
        \mkern-6mu \mathord\rightarrow$}        
\def\dvec#1{\vbox{\ialign{##\crcr
        \leftrightarrowfill\crcr\noalign{\kern-1pt\nointerlineskip}
        $\hfil\displaystyle{#1}\hfil$\crcr}}}           

\def\-{\hphantom{-}}
\newcommand{\dd}{\mbox{d}}

\newcommand{\sm}[1]{\mbox{\scriptsize #1}}

\begin{document}

\begin{flushright}
\widetext{DESY-06-213, hep-th/yymmnnn}
\end{flushright}

\title{Conformally Coupled Scalars, Instantons and Vacuum Instability in AdS$_4$}

\date{\today}

\author{Sebastian de Haro}\thanks{Department of Mathematics, King's College, London WC2R 2LS, UK.}
\author{Ioannis Papadimitriou}\thanks{
DESY Theory Group, Notkestrasse 85, D-22603 Hamburg, Germany.}
\author{Anastasios C. Petkou}
\thanks{Department of Physics, University of Crete, 71003 Heraklion, Greece.}

\begin{abstract}

We show that a scalar field conformally coupled to AdS gravity in four dimensions with a quartic self-interaction
can be embedded into M-theory. The holographic effective potential
is exactly calculated, allowing us to study non-perturbatively the stability of AdS$_4$ in
the presence of the conformally coupled scalar. It is shown that there exists a one-parameter family of conformal
scalar boundary conditions for which the boundary theory has an unstable vacuum. In this case,
the bulk theory has instanton solutions that mediate the decay of the AdS$_4$ space. These results match
nicely with the vacuum structure and the existence of instantons in an effective three-dimensional
boundary model.
\end{abstract} \maketitle

{\bf Introduction:}
Quantum mechanical tunneling is an important mechanism for vacuum selection in the huge landscape of string
theory vacua. It is also expected that it plays a role in the early universe, and in fact it can become
dominant in an eternally inflating universe. In both cases tunneling needs to be understood in the presence
of a quantum gravity theory. This is a difficult problem. However, if holography is at work one may hope to
map this into a problem of vacuum decay in a theory without gravity which might be more accessible. In any
case, it is certainly of interest to study holography in the presence of unstable vacua.

As a step in this direction we study here a simple model that arises as a consistent truncation
of M-theory to four dimensions. The model consists of a scalar field conformally coupled to gravity with
a quartic self-interaction. We calculate the exact holographic effective potential of
the dual boundary theory on the sphere and we show that there
exists a one-parameter family of boundary conditions for the
scalar field such that, in a certain range of this parameter, the boundary theory has an unstable vacuum.
In the same parameter range we find instanton solutions whose Lorentzian signature form describes a
bubble of true vacuum  expanding at the speed of light \cite{Coleman:1980aw}.
However, to understand the end-point of the decay it is necessary to go beyond the supergravity approximation and
consider finite-$N$ corrections.
Finally, we argue that an $O(N')$ $\phi^6$ three-dimensional theory
qualitatively reproduces the bulk results.

{\bf Conformally coupled scalars and their embedding in string/M-theory:}
The action of a scalar field conformally coupled to Einstein gravity
with a negative cosmological constant in four dimensions is
\be\label{conformal_action}
S=\half\int\dd^4x\sqrt{g}\left(\frac{-R+2\L}{\k^2}+(\pa_\m\f)^2+\frac{1}{6}R\f^2+\l\f^4\right)
\ee
where $\k^2=8\p G_4$, $\l$ is a dimensionless coupling and the cosmological constant is $\L=-3/l^2$.
An important property of (\ref{conformal_action}) is that all its extrema have constant Ricci scalar $R=-12/l^2$ \cite{IP}.

It was pointed out in \cite{MTZ} that, for the special value $\l=\k^2/6l^2$,
(\ref{conformal_action}) can be obtained via the field redefinition
\be\label{field_redef}
\k\f/\sqrt{6}=\tanh(\k\tilde{\f}/\sqrt{6}),
\quad g_{\m\n}=\cosh^2(\k\tilde{\f}/\sqrt{6})\tilde{g}_{\m\n},
\ee
from an action with a minimally coupled scalar field and the potential
$V(\tilde{\f})=-(3/\k^2l^2)\cosh(\sqrt{2}\k\tilde{\f}/\sqrt{3})$.
The resulting action is a consistent truncation of the $\cn=8$ gauged supergravity action to the diagonal
of the Cartan subgroup $U(1)^4$ of the $SO(8)$ gauge group \cite{Duff:1999gh}. It follows that any solution of
(\ref{conformal_action}) with $\l=\k^2/6l^2$, can be uplifted to a solution of eleven-dimensional supergravity.
The explicit uplift for this particular
one-scalar truncation takes the form \cite{Papadimitriou:2006dr}:
\bea\label{11D_metric}
d\hat{s}_{11}^2&=&4(X+X^{-1})^{-2}\tilde{\D}^{2/3}ds_{4}^2\NO\\
&&+4l^2\tilde{\D}^{-1/3}\left[X^3\left((\cos^2\th+X^{-4}\sin^2\th)d\th^2
\right.\right.\NO\\
&&\left.\left.+\sin^2\th d\f_4^2\right)+X^{-1}\cos^2\th d\Om_5^2 \right],\\
\label{11D_four_form}\hat{F}\sub{4}&=&-16l^{-1}\left(X+X^{-1}\right)^{-4}\times\NO\\
&&\left(2X^2\cos^2\th+X^{-2}(1+2\sin^2\th)\right)\e_{4}\\
&& +16l\sin2\th\left(X+X^{-1}\right)^{-2}X^{-1}\ast_{4}dX\wedge d\th,\NO
\eea
where we have defined $\tilde{\D}=X\cos^2\th+X^{-3}\sin^2\th$ and $X=(1+\k\f/\sqrt{6})^{1/2}/(1-\k\f/\sqrt{6})^{1/2}$.
Note that the metric (\ref{11D_metric}) preserves an $S^5$ and contains
a squashed $S^2$, which becomes totally squashed as $\f\to\sqrt{6}/\k$, or $X\to\infty$.
This signals a breakdown of the supergravity description in this limit.

{\bf Boundary conditions and the holographic
effective action:}
A scalar field in (Euclidean) AdS$_4$ with radius $l$ and in the upper half plane coordinates
has the asymptotic behavior
\be\label{scalarasympt}
\f\sim z^{\D_-}\left(\f_-(\vec{z})+\cdots\right)+
z^{\D_+}\left(\f_+(\vec{z})+\cdots\right),
\ee
as $z$ approaches the conformal boundary at $z=0$. The parameters $\D_\pm$, where $\D_+\geq \D_-$,
are related to the mass of the scalar by $m^2l^2=\D(\D-3)$ and $\f_\pm(\vec{z})$ are
arbitrary functions of the transverse coordinates. It is known that such a scalar can be consistently quantized in AdS$_4$
either with  Dirichlet, $\d\f_-(\vec{z})=0$, or with Neumann, $\d\f_+(\vec{z})=0$, boundary conditions if the mass squared is
in the range $-9/4<m^2l^2<-5/4$ \cite{Breitenlohner:1982jf}.
This is the case for the scalar $\phi$ in (\ref{conformal_action}) since the scalar curvature $R$ is
constant and acts as a mass term with $m^2l^2=-2$. Quantizing $\phi$ with Neumann boundary conditions one concludes that the
dual boundary theory
has an operator with dimension $\D_-=1$ \cite{Klebanov:1999tb}.
This is consistent with the M-theory embedding discussed above, where
the scalar $\tilde{\f}$ in (\ref{field_redef}), and more generally all the scalars of
the $SL(8,\mathbb{R})/SO(8)$ submanifold of the scalar manifold of $\cn=8$
gauged supergravity, is dual to a dimension one operator whose VEV
parameterizes a certain direction of the Coulomb branch of the $\cn=8$
SCFT on the worldvolume of coincident M2-branes.

In AdS/CFT the (generically non-local) relationship between the functions $\phi_-$ and $\phi_+$ in the asymptotic expansion
(\ref{scalarasympt}), which is imposed by the requirement of regularity of the exact solution, determines the effective action of the boundary
theory. In particular, the effective action of the dual boundary theory  is given by the on-shell value of the renormalized bulk action $S[\phi_-]$,
taken as a functional of $\phi_-$, by the
relation $\Gamma_{\sm{eff}}[-\phi_-]=S[\phi_-]$. This follows from the fact that $S[\phi_-]$ is minus the generating functional for
connected correlation functions of the boundary operator with dimension $\Delta_+$ and its Legendre transform gives the
corresponding generating functional for the boundary operator with dimension $\Delta_-$. It follows that the relationship
$
\delta\Gamma_{\sm{eff}}[\phi_-]/\delta\phi_- =-\delta S[\phi_-]/\delta\phi_-=0
$
determines the vacuum structure of the dual boundary
theory.

Starting with solutions satisfying Neumann boundary conditions with vanishing source $J(\vec z)$ for the dual operator, i.e. $\phi_+(\vec z)\equiv J(\vec z)=0$,
one can modify the boundary conditions while preserving the (bosonic) asymptotic
symmetry group of AdS$_4$. There is a one-parameter family of such deformations
\be\label{bc0}
\phi_+(\vec z)=-l\a\,\phi^2_-(\vec z)~.
\ee
These mixed boundary conditions interpolate between Neumann  ($\a=0$) and Dirichlet ($\a=\infty$). For generic $\a$, the
new boundary condition will not be a stationary point of (\ref{conformal_action}) but it can be enforced by adding a
boundary term to the action which we will determine. Eventually we will be considering solutions with vanishing stress-energy
tensor and for that we renormalize to zero the on-shell contribution of the gravity part of the action. The matter part of the
action includes also a generalized Gibbons-Hawking term (see \cite{IP} for details). The boundary term that enforces (\ref{bc0}) is
\be\label{b_term}
S_{\sm{bdy}}=-\frac{l^3\a}{3}\int\dd^3z\f_-^3(\vec{z})~.
\ee
In the context of the AdS/CFT correspondence the addition of the boundary term (\ref{b_term})
corresponds to a marginal triple-trace deformation of the dual CFT
\cite{Hertog:2004dr}, completely breaking supersymmetry. The boundary condition (\ref{bc0}) has
been studied in the context of `designer gravity' \cite{Hertog:2004ns}, where various black hole solutions \cite{Hertog:2004dr}
as well as gravitational solitons and cosmological big bang/crunch geometries \cite{Hertog:2004rz} satisfying
these boundary conditions were numerically constructed. An exact Poincar\'e domain wall solution satisfying the boundary
condition (\ref{bc0}) was found and uplifted to eleven dimensions in \cite{Papadimitriou:2006dr}.

The equations of motion following from (\ref{conformal_action}) determine in principle the non-local relation between the two modes,
$\f_\pm(\vec{z})$, and hence the holographic effective action for the VEV of the dual dimension one operator in the CFT deformed by the marginal
deformation (\ref{b_term}). This effective action can be computed in a derivative expansion away from the vanishing VEV point. On
a nearly flat boundary one finds that up to two derivatives \cite{IP}
\bea\label{eff_action}
\G_{\sm{eff}}[\f_-]&=&\frac{1}{6\sqrt{\l}}\int d^3x\sqrt{g\sub{0}} \left(\rule{0pt}{15pt}\f_-^{-1}\pa_i\f_-\pa^i\f_-\right.\NO\\
&&\left.+\frac12 R[g\sub{0}]\f_-+2\sqrt{\l}(\sqrt{\l}-\a)\f_-^3\right),
\eea
where $g\sub{0}_{ij}$ is the boundary metric. Moreover,  the exact holographic effective potential for $R\geq 0$  is \cite{IP}
\be\label{eff_potential}
V_{\l,\a}(\f_-)=\frac{1}{3\l}\left[\left(\frac{R}{6}+\l\f_-^2\right)^{3/2}-\a\l\f_-^3
-\left(\frac{R}{6}\right)^{3/2}\right],
\ee
where the additive constant has been fixed by requiring that the trivial vacuum at $\f_-=0$
has zero energy. Redefining $\f_-=\vf^2$ and taking  $\a\to\sqrt{\l}$  the two-derivative holographic effective
action  (\ref{eff_action}) takes the form
\bea\label{lim_eff_action}
\G_{\sm{eff}}[\vf]&=&\frac{4}{3\sqrt{\l}}\int d^3x\sqrt{g\sub{0}} \left(\rule{0pt}{15pt}\frac{1}{2}\pa_i\vf\pa^i\vf\right.\NO\\
&&\left.+\frac{1}{16} R[g\sub{0}]\vf^2+\frac{1}{8}\m\vf^6\right)+\co\left(\m^2\right),
\eea
where $\m\equiv \l-\a^2$. This agrees with the classically conformally invariant toy-model actions used in \cite{dH&P,Hertog:2004rz}.

Since the system (\ref{conformal_action}) can be embedded into eleven-dimensional
supergravity, the holographically dual field theory is (a sector of)
the $\cn=8$ interacting SCFT in the large-$N$ limit \cite{Seiberg:1997ax}.
In the abelian case, $N=1$, this theory can be obtained by compactifying
$\cn=4$ super Yang-Mills in four dimensions on a circle in the limit of zero radius,
and it is also believed to be the infrared fixed point of $\cn=8$ super Yang-Mills
in three dimensions. However little is known about this theory, which makes any test of the AdS/CFT correspondence very difficult.
Nevertheless, we will argue below that the bulk results (\ref{eff_action}), (\ref{eff_potential}) and (\ref{lim_eff_action}) can be
qualitatively reproduced by a certain three-dimensional $O(N')$ model, where $N'$ is not related to the number of $M2$-branes. This not only
implies some connection between the $\cn=8$ SCFT and the $O(N')$ model, as a consequence of
the AdS/CFT correspondence, but it is also a step towards finding the holographic dual
of $O(N')$ models in three dimensions \cite{Klebanov:2002ja}.

{\bf Scalar instantons:}
The equations of motion following from the action (\ref{conformal_action})
admit non-trivial solutions with vanishing stress tensor.
The condition that the stress tensor vanishes reduces to a linear equation for the scalar field, namely
$(\nabla_\m\nabla_\n-\frac{1}{4} g_{\m\n}\square_g)\f^{-1}=0$ \cite{IP},
which admits non-trivial solutions provided the metric is that of exact AdS$_4$. In the upper half plane
coordinates the most general solution of the equations of motion with vanishing bulk stress tensor takes the form
\be\label{instanton_uhp}
\f=\frac{2}{l\sqrt{|\l|}}\left(\frac{bz}{-{\rm sgn}(\l)b^2
+(z+a)^2+(\vec{z}-\vec{z}_0)^2}\right),
\ee
where $a,b,z_0^i$, $i=1,2,3$, are arbitrary constants. For $\l>0$,
this Euclidean solution is non-singular provided
\be\label{condition}
a>b\geq 0.
\ee
For $\lambda<0$ this solution was studied  in \cite{dH&P} ignoring its backreaction on the geometry.
We have now shown that in fact there is no backreaction and
(\ref{instanton_uhp}) together with the AdS$_4$ metric
is an exact solution of the coupled equations of motion.
Expanding the solution (\ref{instanton_uhp}) asymptotically near the conformal boundary
we get precisely the expansion (\ref{scalarasympt}), where $\f_\pm$ satisfy (\ref{bc0}) with
$\a\equiv\sqrt{|\l|}a/b$ and
\be\label{b_instanton}
\f_-=\frac{2}{l\sqrt{|\l|}}\left(\frac{b}{-{\rm sgn}(\l)b^2
+a^2+(\vec{z}-\vec{z}_0)^2}\right).
\ee
Notice that the parameter $\a$ is {\em not} a modulus of the solution but rather labels different boundary conditions.
Therefore, it corresponds to the deformation parameter of the dual theory and different
values of $\a$ correspond to saddle points of different theories. The
remaining parameters in (\ref{instanton_uhp}) do correspond to moduli of
the solution and they parameterize a four-dimensional hyperbolic space
$\widetilde{\mathbb{H}}_4$ of radius $\tilde{l}=\sqrt{\a^2-\l}$, which is also the
moduli space of instantons in ${\mathbb R}^3$. This can be seen by parameterizing
the solution (\ref{instanton_uhp}) in a manifestly $O(5,1)$ covariant form as in \cite{fubini},
$\f^{-1}=h_My^M$, where $h,y$ are vectors in a 6-dimensional embedding space. It can then be
shown \cite{IP} that the solution is determined by the 5-dimensional part of $h$, which satisfies
$h^2=-\frac{|\l|}{b^2}(a^2-b^2{\mbox{sgn}}(\l))$ and hence, given (\ref{condition}),
it parametrizes an $\widetilde{\mathbb{H}}_4$ also for $\l>0$.
Note that the condition (\ref{condition}) for the solution to be
non-singular is the necessary and sufficient condition for the radius of the
moduli space to be well defined and also for the potential (\ref{eff_potential})
to be unbounded from below. Remarkably, the VEV (\ref{b_instanton}) is an exact extremum
of the two-derivative holographic effective action (\ref{lim_eff_action}). 

{\bf Vacuum decay:}
The unboundedness of the boundary effective potential for $\a>\k/\sqrt{6}l$ is the holographic
image of an AdS$_4$ instability towards its spontaneous
`dressing' by a non-zero scalar field when the boundary condition
(\ref{bc0}) is imposed with $\a>\k/\sqrt{6}l$.
This non-perturbative instability does not contradict any positivity
theorem. Although the positivity theorems \cite{Breitenlohner:1982jf,
Breitenlohner:1982bm,Gibbons:1983aq,Townsend:1984iu} do apply
when the supersymmetric ($\a=\k/\sqrt{6}l$ \cite{IP}) boundary conditions are imposed
on the scalar field, they do not apply for a generic value of the deformation
parameter $\a$ in (\ref{b_term}).
The question of stability of AdS with such boundary conditions has been addressed in
the context of `designer gravity'
\cite{Hertog:2004ns,Hertog:2005hm}, where it has been suggested that AdS
with boundary conditions on the scalar defined by the boundary term
$S_\cw=-\int d^3z\cw(\f_-)$, is stable provided the function $\cw(\f_-)$ has a global minimum.
This is clearly not the case for the deformation (\ref{b_term}) and so
these positivity theorems do not apply either.

\begin{figure}
\scalebox{0.8}{\rotatebox{-0}{\includegraphics{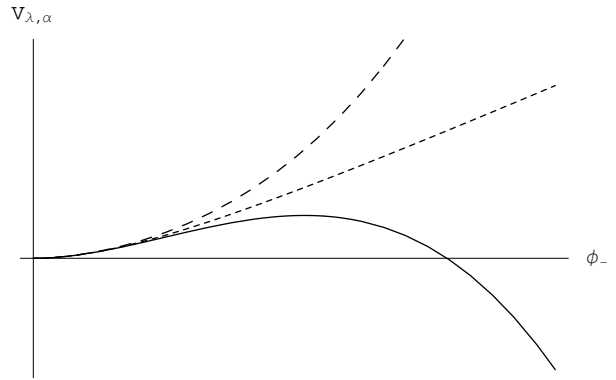}}}
\caption{Plot of the potential (\ref{eff_potential}) on $S^3$ for $\a<\k/\sqrt{6}l$
(long dashes), $\a=\k/\sqrt{6}l$ (short dashes), and $\a>\k/\sqrt{6}l$.}
\label{fig}
\end{figure}

The physical meaning of the instanton solutions (\ref{instanton_uhp}) becomes clear by taking the boundary to be $S^3$.
In this case the potential (\ref{eff_potential}) has a global minimum at $\f_-=0$ for $\a\leq \sqrt{\l}$ (note that
$\f_-\geq 0$) which turns into a local minimum for $\a>\sqrt{\l}$ separated by a potential barrier from the instability region
at $\phi_- \to \infty$ (Fig. \ref{fig}). The scalar field can tunnel
from the local minimum at $\f_-=0$ to the instability region.
In terms of the bulk scalar field the instability region is reached for $\f\to\sqrt{6}/\k$,
or $\tilde{\f}\to\pm\infty$. We have seen that this is precisely the limit where an $S^2$ in the
uplifted eleven-dimensional metric (\ref{11D_metric}) gets totally squashed giving rise to a
singularity. To understand the true vacuum of the theory for $\a>\k/\sqrt{6}l$, if there is one,
one needs to go beyond supergravity to include finite-$N$ corrections. This would modify
the holographic potential (\ref{eff_potential}) and at the same time would resolve the geometric
singularity from the squashed $S^2$.

The physical picture of the decay process can be understood in terms
of bubble nucleation \`a la Coleman and De Luccia \cite{Coleman:1980aw}.
Namely, the Lorentzian version of the instanton solution (\ref{instanton_uhp})
describes a `half-bubble' centered on the boundary which is expanding or collapsing at the speed of
light towards the bulk. The equatorial plane of the bubble describes an
expanding or collapsing bubble in the dual field theory. Outside and far away from the bubble the space is AdS$_4$
corresponding to the local minimum at $\f_-=0$ of the potential (\ref{eff_potential}).
Inside the bubble is the `true vacuum' which cannot be understood
in the supergravity approximation. The tunneling probability is proportional to the exponential of the
on-shell action computed with the boundary conditions defined by
(\ref{b_term}) and evaluated on the instanton solution (\ref{instanton_uhp}),
$\cp\propto \exp{(-\left.\G_{\sm{eff}}\right|_{\sm{inst}})}$. Evaluating this gives \cite{IP}
\be\label{decay_rate}
\left.\G_{\sm{eff}}\right|_{\sm{inst}}=\frac{4\pi^2l^2}{\k^2}\left(\frac{1}{\sqrt{1-\k^2/6l^2\a^2}}-1\right).
\ee
Note that the deformation parameter $\a$ drives the theory from the regime of marginal
stability at $\a=\k/\sqrt{6}l$ to total instability at $\a\to\infty$.
In the global coordinates of $\mathbb{H}_4$ where the boundary is $S^3$, the instanton
solution (\ref{instanton_uhp}) depends only on the $\mathbb{H}_4$ radius. In particular,
$\f_-$ is constant in these coordinates. This positive constant solution corresponds
precisely to the local maximum of the holographic potential (\ref{eff_potential}) on $S^3$,
which again only exists for $\a>\k/\sqrt{6}l$. Evaluating the potential at this maximum
and multiplying with the volume of $S^3$ we get exactly (\ref{decay_rate}). This confirms
that the tunneling probability is indeed proportional to the exponential of minus
the height of the potential barrier and justifies our claim that the instantons mediate
the tunneling of the local minimum at $\f_-=0$ to the instability region for $\a>\k/\sqrt{6}l$.

{\bf An effective boundary theory:}

There is some indication that the holographic results (\ref{eff_potential}) and
(\ref{decay_rate}) could be reproduced from an $O(N')$ model in three dimensions with a
$g(\vec\phi^2)^3/6$ interaction. Both the form (\ref{lim_eff_action})
of the holographic effective action in the double scaling limit and
the existence of instantons in certain $O(N')$ models in three dimensions are suggestive of
such a connection.

To illustrate this point we consider the action
\be
\label{phi6_generic}
I^g=\cc\int d^3\vec{x}\left(\frac{1}{2}\partial_i\phi^a\partial^i\phi^a +\frac{g}{6N'^2}(\phi^a\phi^a)^3\right),
\ee
where $a=1,2,\ldots,N'$ and $\cc$ is an $N'$-independent constant. For $g<0$ this model has instanton configurations given by
\be
\label{phi6_instantons}
\phi^a(\vec{x}) =\left(3N'^2/(-g)\right)^{1/4}(c^a/\sqrt{b})\left(b^2+(\vec{x}-\vec{x}_0)^2\right)
^{-\frac12}~,
\ee
where $c^ac^a=b^2$ and $b$ is an arbitrary constant. On $S^3$ the classical
potential for the $O(N')$-singlet operator $\s\equiv \f^a\f^a/N'$ becomes
$V_g(\s)=\cc\left(R\s/16+g\s^3/6\right)$. With the identifications
$\cc=4/3\sqrt{\l}$, $g=3\sqrt{\l}(\sqrt{\l}-\a)/2$ and $\s=\f_-$, this potential
coincides with the holographic potential (\ref{eff_potential}) in the limit
of small curvature. In particular, the instability region of (\ref{eff_potential})
for $\a>\sqrt{\l}$ is mapped precisely to the instability region of
the $O(N')$ model for $g<0$. It would be very interesting if a full quantum treatment
of this $O(N')$ model reproduced the full holographic potential (\ref{eff_potential})
in the large-$N'$ limit \cite{toappear}. Evaluating the action (\ref{phi6_generic}) on the
instanton solution (\ref{phi6_instantons})
we obtain $N'^{-1}I_{\sm{inst}}^{g}=(\sqrt{2}\p^2/3\l)(\a/\sqrt{\l}-1)^{-1/2}$,
which precisely agrees with the tunneling probability (\ref{decay_rate}) in the approximation
$\a/\sqrt{\l}\approx 1$.

\centerline{\bf Acknowledgements} 

I.P. thanks the Aspen Center for Physics for hospitality during the
workshop on `Recent Advances in Black Hole Physics in String Theory',
as well as the Albert Einstein Institute at Potsdam and the University of Crete, where part of
this work was done. A.C.P. wishes to thank T.~Tomaras for very useful discussions. A.C.P. is partially supported by the
European RTN ``Superstrings'' and the
Greek Research Program ``PYTHAGORAS II''.

\end{document}